\documentclass[aps,prl,superscriptaddress,showpacs,floatfix,twocolumn]{revtex4}

\usepackage{amsmath}
\usepackage{graphicx}

\begin{document}


\title{Particle-species dependent modification of jet-induced correlations 
in Au+Au collisions at $\sqrt{s_{NN}}$ = 200 GeV}

\newcommand{\abilene}{Abilene Christian University, Abilene, TX 79699, USA}
\newcommand{\banaras}{Department of Physics, Banaras Hindu University, Varanasi 221005, India}
\newcommand{\bnl}{Brookhaven National Laboratory, Upton, NY 11973-5000, USA}
\newcommand{\caucr}{University of California - Riverside, Riverside, CA 92521, USA}
\newcommand{\cns}{Center for Nuclear Study, Graduate School of Science, University of Tokyo, 7-3-1 Hongo, Bunkyo, Tokyo 113-0033, Japan}
\newcommand{\colorado}{University of Colorado, Boulder, CO 80309, USA}
\newcommand{\columbia}{Columbia University, New York, NY 10027 and Nevis Laboratories, Irvington, NY 10533, USA}
\newcommand{\dapnia}{Dapnia, CEA Saclay, F-91191, Gif-sur-Yvette, France}
\newcommand{\debrecen}{Debrecen University, H-4010 Debrecen, Egyetem t{\'e}r 1, Hungary}
\newcommand{\elte}{ELTE, E{\"o}tv{\"o}s Lor{\'a}nd University, H - 1117 Budapest, P{\'a}zm{\'a}ny P. s. 1/A, Hungary}
\newcommand{\fsu}{Florida State University, Tallahassee, FL 32306, USA}
\newcommand{\gsu}{Georgia State University, Atlanta, GA 30303, USA}
\newcommand{\hiroshima}{Hiroshima University, Kagamiyama, Higashi-Hiroshima 739-8526, Japan}
\newcommand{\ihepprot}{IHEP Protvino, State Research Center of Russian Federation, Institute for High Energy Physics, Protvino, 142281, Russia}
\newcommand{\illuiuc}{University of Illinois at Urbana-Champaign, Urbana, IL 61801, USA}
\newcommand{\isu}{Iowa State University, Ames, IA 50011, USA}
\newcommand{\jinrdubna}{Joint Institute for Nuclear Research, 141980 Dubna, Moscow Region, Russia}
\newcommand{\kaeri}{KAERI, Cyclotron Application Laboratory, Seoul, Korea}
\newcommand{\kek}{KEK, High Energy Accelerator Research Organization, Tsukuba, Ibaraki 305-0801, Japan}
\newcommand{\kfki}{KFKI Research Institute for Particle and Nuclear Physics of the Hungarian Academy of Sciences (MTA KFKI RMKI), H-1525 Budapest 114, POBox 49, Budapest, Hungary}
\newcommand{\korea}{Korea University, Seoul, 136-701, Korea}
\newcommand{\kurchatov}{Russian Research Center ``Kurchatov Institute", Moscow, Russia}
\newcommand{\kyoto}{Kyoto University, Kyoto 606-8502, Japan}
\newcommand{\labllr}{Laboratoire Leprince-Ringuet, Ecole Polytechnique, CNRS-IN2P3, Route de Saclay, F-91128, Palaiseau, France}
\newcommand{\lawllnl}{Lawrence Livermore National Laboratory, Livermore, CA 94550, USA}
\newcommand{\losalamos}{Los Alamos National Laboratory, Los Alamos, NM 87545, USA}
\newcommand{\lpc}{LPC, Universit{\'e} Blaise Pascal, CNRS-IN2P3, Clermont-Fd, 63177 Aubiere Cedex, France}
\newcommand{\lund}{Department of Physics, Lund University, Box 118, SE-221 00 Lund, Sweden}
\newcommand{\muenster}{Institut f\"ur Kernphysik, University of Muenster, D-48149 Muenster, Germany}
\newcommand{\myongji}{Myongji University, Yongin, Kyonggido 449-728, Korea}
\newcommand{\nagasaki}{Nagasaki Institute of Applied Science, Nagasaki-shi, Nagasaki 851-0193, Japan}
\newcommand{\newmex}{University of New Mexico, Albuquerque, NM 87131, USA }
\newcommand{\nmsu}{New Mexico State University, Las Cruces, NM 88003, USA}
\newcommand{\ornl}{Oak Ridge National Laboratory, Oak Ridge, TN 37831, USA}
\newcommand{\orsay}{IPN-Orsay, Universite Paris Sud, CNRS-IN2P3, BP1, F-91406, Orsay, France}
\newcommand{\pnpi}{PNPI, Petersburg Nuclear Physics Institute, Gatchina, Leningrad region, 188300, Russia}
\newcommand{\riken}{RIKEN, The Institute of Physical and Chemical Research, Wako, Saitama 351-0198, Japan}
\newcommand{\rikjrbrc}{RIKEN BNL Research Center, Brookhaven National Laboratory, Upton, NY 11973-5000, USA}
\newcommand{\rikkyo}{Physics Department, Rikkyo University, 3-34-1 Nishi-Ikebukuro, Toshima, Tokyo 171-8501, Japan}
\newcommand{\saispbstu}{Saint Petersburg State Polytechnic University, St. Petersburg, Russia}
\newcommand{\saopaulo}{Universidade de S{\~a}o Paulo, Instituto de F\'{\i}sica, Caixa Postal 66318, S{\~a}o Paulo CEP05315-970, Brazil}
\newcommand{\seoulnat}{System Electronics Laboratory, Seoul National University, Seoul, Korea}
\newcommand{\stonybrkc}{Chemistry Department, Stony Brook University, Stony Brook, SUNY, NY 11794-3400, USA}
\newcommand{\stonycrkp}{Department of Physics and Astronomy, Stony Brook University, SUNY, Stony Brook, NY 11794, USA}
\newcommand{\subatech}{SUBATECH (Ecole des Mines de Nantes, CNRS-IN2P3, Universit{\'e} de Nantes) BP 20722 - 44307, Nantes, France}
\newcommand{\tenn}{University of Tennessee, Knoxville, TN 37996, USA}
\newcommand{\titech}{Department of Physics, Tokyo Institute of Technology, Oh-okayama, Meguro, Tokyo 152-8551, Japan}
\newcommand{\tsukuba}{Institute of Physics, University of Tsukuba, Tsukuba, Ibaraki 305, Japan}
\newcommand{\vandy}{Vanderbilt University, Nashville, TN 37235, USA}
\newcommand{\waseda}{Waseda University, Advanced Research Institute for Science and Engineering, 17 Kikui-cho, Shinjuku-ku, Tokyo 162-0044, Japan}
\newcommand{\weizmann}{Weizmann Institute, Rehovot 76100, Israel}
\newcommand{\yonsei}{Yonsei University, IPAP, Seoul 120-749, Korea}
\affiliation{\abilene}
\affiliation{\banaras}
\affiliation{\bnl}
\affiliation{\caucr}
\affiliation{\cns}
\affiliation{\colorado}
\affiliation{\columbia}
\affiliation{\dapnia}
\affiliation{\debrecen}
\affiliation{\elte}
\affiliation{\fsu}
\affiliation{\gsu}
\affiliation{\hiroshima}
\affiliation{\ihepprot}
\affiliation{\illuiuc}
\affiliation{\isu}
\affiliation{\jinrdubna}
\affiliation{\kaeri}
\affiliation{\kek}
\affiliation{\kfki}
\affiliation{\korea}
\affiliation{\kurchatov}
\affiliation{\kyoto}
\affiliation{\labllr}
\affiliation{\lawllnl}
\affiliation{\losalamos}
\affiliation{\lpc}
\affiliation{\lund}
\affiliation{\muenster}
\affiliation{\myongji}
\affiliation{\nagasaki}
\affiliation{\newmex}
\affiliation{\nmsu}
\affiliation{\ornl}
\affiliation{\orsay}
\affiliation{\pnpi}
\affiliation{\riken}
\affiliation{\rikjrbrc}
\affiliation{\rikkyo}
\affiliation{\saispbstu}
\affiliation{\saopaulo}
\affiliation{\seoulnat}
\affiliation{\stonybrkc}
\affiliation{\stonycrkp}
\affiliation{\subatech}
\affiliation{\tenn}
\affiliation{\titech}
\affiliation{\tsukuba}
\affiliation{\vandy}
\affiliation{\waseda}
\affiliation{\weizmann}
\affiliation{\yonsei}
\author{S.~Afanasiev}	\affiliation{\jinrdubna}
\author{C.~Aidala}	\affiliation{\columbia}
\author{N.N.~Ajitanand}	\affiliation{\stonybrkc}
\author{Y.~Akiba}	\affiliation{\riken} \affiliation{\rikjrbrc}
\author{J.~Alexander}	\affiliation{\stonybrkc}
\author{A.~Al-Jamel}	\affiliation{\nmsu}
\author{K.~Aoki}	\affiliation{\kyoto} \affiliation{\riken}
\author{L.~Aphecetche}	\affiliation{\subatech}
\author{R.~Armendariz}	\affiliation{\nmsu}
\author{S.H.~Aronson}	\affiliation{\bnl}
% ...
\collaboration{PHENIX Collaboration} \noaffiliation

\date{\today}

\begin{abstract}
We report PHENIX measurements of the correlation of a trigger hadron, at
intermediate transverse momentum ($2.5<p_{\rm T,trig}<4$ GeV/$c$), with associated
mesons or baryons, at lower $p_{\rm T,assoc}$, in Au+Au collisions at 
$\sqrt{s_{NN}}=200$ GeV. The jet correlations, for both baryons and mesons, show 
similar shape alterations as a function of centrality, characteristic of strong 
modification of the away-side jet. The ratio of jet-associated baryons to mesons 
for this jet, increases with centrality and $p_{\rm T,assoc}$ and, in the most 
central collisions, reaches a value similar to that for inclusive measurements. 
This trend is incompatible with in-vacuum fragmentation, but could be due to 
jet-like contributions from correlated soft partons which recombine upon 
hadronization.
%
%

\end{abstract}

\maketitle

	Recent measurements at the Relativistic Heavy Ion 
Collider (RHIC) have indicated the creation of a new state of matter in 
heavy-ion collisions~\cite{Adcox:2004mh}.
%
The ``soft" or small momentum transfer processes leading to the formation 
of this collision medium are sometimes accompanied by hard parton-parton 
scatterings. These scattered partons interact strongly with the medium and 
lose energy as they propagate through it, before fragmenting  into 
jets~\cite{Gyulassy:2003mc,Baier:1996kr}. 
This can lead to strong modification of both the yield and 
the angular correlation patterns of jets~\cite{Adler:2005ee,Adare:2007vu}. 
Therefore, the study of jets can provide invaluable insights into the properties 
of the new state of matter.
 
 	Parton energy loss in the nuclear collision medium 
\cite{Gyulassy:2003mc,Baier:1996kr} has 
been associated with the observation that the single particle yields of 
mesons ($M$) are significantly suppressed in Au+Au collisions, when compared 
to the yields in p+p collisions scaled by the number of binary nucleon-nucleon
collisions~\cite{Adcox:2001jp,Adams:2003kv}. This suppression factor 
is $R^M_{\rm AA} \sim 0.2$, for transverse momentum $p_{\rm T} \agt 4$~GeV/$c$ 
(in the absence of suppression, $R_{\rm AA} = 1.0$). In contrast to meson behavior, 
a general pattern of baryon ($B$) enhancement 
(for intermediate $p_{\rm T} \sim 2-5$~GeV/$c$) relative to mesons has been observed 
in central Au+Au collisions at RHIC~\cite{Adcox:2001mf,Adler:2002uv}. 
This is dramatized by a strikingly large proton to pion ratio which is about 
three times larger than in p+p collisions~\cite{Adcox:2004mh}. In fact, 
there is no suppression for baryons for $p_{\rm T} \sim 1.5-4$~GeV/$c$ 
(i.e. ($R^B_{\rm AA} \sim 1.0$))~\cite{Adler:2003pzl}, compared to the very strong 
suppression for mesons. 

	Quark recombination~\cite{Voloshin:2002wa,Greco:2003mm,Fries:2003kq}
has been used to explain the enhancement of baryon emission in the intermediate
$p_{\rm T}$ range. Such models also provide an explanation for the observed dependence 
of the elliptic flow on hadron species in terms of the ``universal" elliptic flow of 
constituent quarks~\cite{Adare:2006ti,Lacey:2006pn}. By contrast, results from jet-induced 
hadron correlation measurements~\cite{Adler:2004zd} rule out simple models 
which only take account of jet fragmentation and the recombination of thermal 
quarks in a flowing medium. The dichotomy between these two sets of observations 
is currently an unresolved issue at RHIC.

	Full suppression of the away-side jet in Au+Au collisions
has been reported~\cite{Adler:2002ct}. Recently, relative azimuthal angle ($\Delta\phi$) 
correlation measurements of the away-side jet partner hadrons 
at lower momentum have been found to be significantly 
modified~\cite{Adler:2005ee,Adare:2007vu}. 
Indeed, these distributions show local minima at $\Delta\phi = \pi$ which 
contrasts with the characteristic jet peak observed in $p+p$ collisions.
This modification has been linked to strong parton-medium 
interactions~\cite{Adare:2006nr,Adare:2007vu}.  

A crucial question is whether or not such interactions could also induce 
correlations between soft partons (comprising the medium) which could then recombine 
to form jet-like fragments. To this end, we use measurements of $\Delta\phi$ correlation functions  
to make detailed investigations of the distributions and conditional yields of jet 
associated baryons and mesons. The study is made as a function of collision centrality 
and partner $p_{\rm T}$, for the trigger hadron selection $2.5< p_{\rm T,trig} <4.0$~GeV/$c$.

	The hadron yield from jet fragmentation is relatively small in the 
intermediate $p_{\rm T}$ range, and recombination of uncorrelated soft 
partons cannot produce jet-like correlations. Thus, very little away-side 
jet correlation might be expected for associated baryons and mesons.
By contrast, it has been argued~\cite{Fries:2004hd} 
that energy loss, by a hard scattered parton propagating through the 
collision medium, can induce two-body correlations between soft 
partons in a region surrounding the hard parton's trajectory~\cite{new_calc}. 
Soft partons from this region could then recombine 
into hadrons which not only correlate with each other, but also with the direction of 
the hard scattered parton. The process of recombination would also amplify these 
jet-like correlations for baryon creation compared to that for mesons and hence, 
result in particle ratios different from the in-vacuum fragmentation values. 

	Au+Au data (at $\sqrt{s_{NN}}$=200 GeV) was recorded during 2004 with 
the PHENIX detector~\cite{Adcox:2003zm}. Collision centrality was determined 
with the beam-beam counters (BBC) and zero degree 
calorimeters~\cite{Adcox:2003zm}. Charged particle tracking, identification, 
and momentum reconstruction in the central rapidity region ($|\eta|\leq 0.35$) 
was provided by two drift chambers, each with an azimuthal coverage 
$\Delta\varphi=\pi/2$, and two layers of multi-wire proportional chambers with 
pad readout (PC1 and PC3). To reject most background from albedo, conversions, 
and decays, a confirming hit was required within a 2$\sigma$ matching window in 
PC3~\cite{Adcox:2001jp}.

	Charged particles were identified via time-of-flight measurement 
with the time-of-flight (TOF) and lead scintillator (PbSc) detectors.
The TOF covers $\Delta\varphi=\pi/4$ with good timing
resolution $\simeq$ 120 ps (see Ref.~\cite{Adler:2004zd,Adare:2006nn}; 
the PbSc as used here, covers a larger solid angle 
($\Delta\varphi=3\pi/4$) with a modest timing resolution 
of 400 ps. The time-of-flight measurements were used in conjunction 
with the measured momentum and flight-path length, to generate a 
mass-squared ($m^2$) distribution~\cite{Adler:2003cb} for charged particle 
identification. A cut about the baryon ($\bar{p}, p$) peak in the $m^2$ 
distribution was used to distinguish baryons and 
mesons ($\pi^{\pm}, K^{\pm}$). 
The kaon contamination of the baryon sample is $\alt 3\%$ for the highest 
associated $p_{\rm T}$ bin used ($1.6<p_{\rm T,assoc}^{M,B}<2$ GeV/$c$).
We generated area normalized two-particle correlation functions, in relative 
azimuthal angle $C\left(\Delta\phi\right)$, as the ratio of
a foreground distribution $N_{\rm cor}(\Delta\phi)$, constructed with
correlated particle pairs from the same event,
and a background distribution $N_{\rm mix}(\Delta\phi)$, for pairs obtained by mixing
particles from different events having similar collision vertex and 
centrality~\cite{Ajitanand2005,Adler:2005ee};

\begin{equation}
C\left(\Delta\phi\right) = \frac{N_{\rm cor}\left( \Delta\phi
\right)} {N_{\rm mix}\left( \Delta\phi \right)}
\frac{\int d\Delta\phi N_{\rm mix}(\Delta\phi)}{\int d\Delta\phi N_{\rm cor}(\Delta\phi)}. 
\label{eq:cf}
\end{equation}

\begin{figure}[t]
\includegraphics[width=1.0\linewidth]{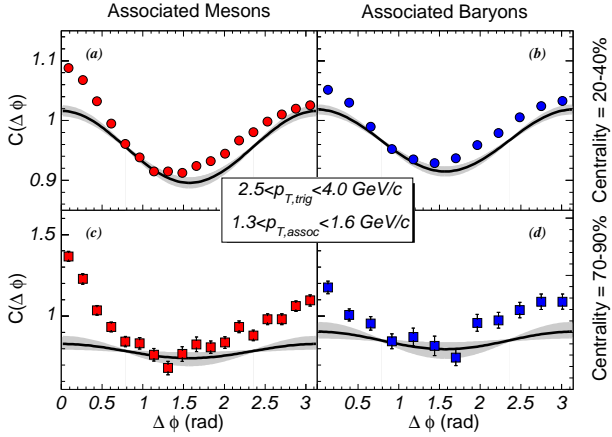}  
\caption{(Color online) Correlation functions for associated partner mesons (a) and (c) and 
         baryons (b) and (d) ($1.3< p_{\rm T} <1.6$~GeV/$c$) per trigger hadron 
         ($2.5< p_{\rm T} <4.0$~GeV/$c$), for centrality selections of 20-40\% (top panels) and 
         70-90\% (bottom panels). The curves indicate elliptic flow contributions (see text).
\label{fig:cf}}
\end{figure}

	Representative examples of the correlation functions, so obtained for associated mesons 
and baryons ($1.3< p_{\rm T,assoc}^{M,B} <1.6$ GeV/$c$) per trigger 
hadron ($2.5< p_{\rm T,trig} <4.0$ GeV/$c$), 
are shown for two centrality selections in Fig~\ref{fig:cf}. They indicate an asymmetry 
characteristic of (di)jet pair correlations (J($\Delta\phi$)) and 
an anisotropy signaling elliptic flow 
($H(\Delta\phi)=1 + 2 (v_{2}^{\rm trig}\times v_{2}^{B,M}) cos(2\Delta\phi)$)
with amplitude $v_{2}^{\rm trig}\times v_{2}^{M,B}$~\cite{Adler:2005ee};
The observed correlation functions for associated partner mesons are  
more asymmetric than those for associated partner baryons, indicating  
that the jet signal is stronger for hadron-meson correlations. However, 
clear separation of the jet and flow correlations 
is required for further study.

	Reliable extraction of J($\Delta \phi$) from $C(\Delta \phi$) 
can be achieved if the normalization b$_{o}$, and 
$v_{2}^{\rm trig}\times v_{2}^{B,M}$ is known~\cite{Ajitanand2005}. 
Values for $v_2^{\rm trig}$ and $v_2^{M,B}$ were 
obtained via measurements of the single particle distributions relative to 
the reaction plane, determined in the BBC's~\cite{Adare:2006ti,Lacey:2006pn}. 
The large (pseudo)rapidity separation ($\Delta \left|\eta \right|> 2.75$), 
between each BBC and the PHENIX central arms, minimizes any non-flow
contributions to these $v_2$ values~\cite{Jia:2006sb}.

To fix the value of b$_{o}$ we followed the 
procedure in Refs.~\cite{Adler:2005ee,Ajitanand2005} and assumed that J($\Delta \phi$) 
has zero yield at some minimum  $\Delta \phi _{\rm min}$ (ZYAM). 
That is, the elliptic flow contributions are required to coincide 
with C($\Delta \phi$) at $\Delta \phi _{\rm min}$. 
Good precision for $\Delta \phi _{\rm min}$ was achieved via a  
fit to the correlation function; the systematic error on the magnitude of 
the integrated jet-function J($\Delta \phi$), due to the ZYAM procedure  
is estimated to be $\alt 3$\%.
The solid lines in Fig.~\ref{fig:cf} show examples of the ZYAM 
normalized elliptic flow ($v_2$) contributions.
The gray bands represent systematic errors on the $v_2$ amplitudes ($\sim 6\%$ for central 
and mid-central events, and $\sim$ 40\% for peripheral events.) 
primarily due to an uncertainty in the reaction plane resolution~\cite{Adler:2005ee}.
%
\begin{figure}[t]
\includegraphics[width=1.0\linewidth]{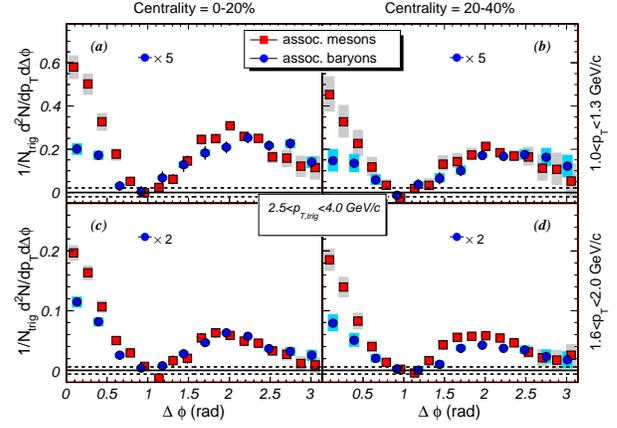}  
\caption{(Color online) Jet-pair distributions for associated mesons (squares) and baryons (circles) 
for $1<p_{\rm T,assoc}<1.3$ GeV/$c$ (top panels) and ($1.6<p_{\rm T,assoc}<2.0$ (bottom panels).
Results are for a hadron trigger ($2.5< p_{\rm T} <4.0$~GeV/$c$) and centrality selections 
of 0-20\% and 20-40\%.
\label{fig:jf}}
\end{figure}

	The associated meson and baryon jet distributions $\frac{1}{N_{\rm trig}}\frac{d^2N}{dp_{\rm T}d\Delta\phi}$
are shown in Fig. \ref{fig:jf} for two associated $p_{\rm T}$ bins 
and for the centralities 0-20\% and 20-40\%. The shaded error bars indicate 
the systematic error related to $v_2$ subtraction. The associated baryon jet 
pair distributions are multiplied by the indicated factors to 
facilitate a shape comparison with the distributions for mesons.

\begin{figure}[t]
\includegraphics[width=1.0\linewidth]{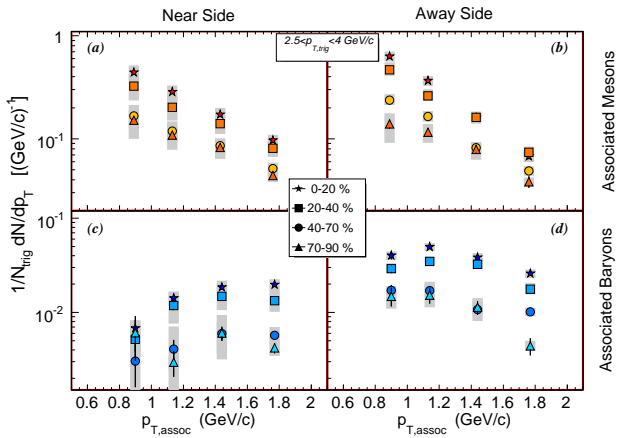}  
\caption{(Color online) conditional jet yields for associated mesons (top panels)
          and baryons (bottom panels) for near- (left panels) 
          and away-side (right panels) jets, as a function of associated particle $p_{\rm T}$
          and collision centrality. 
\label{fig:yld}}
\end{figure}

	Figure \ref{fig:jf} shows that the correlation strength of the 
near side jet ($\Delta \phi \leq \Delta \phi_{\rm min}$, NS) is substantially weaker for 
associated baryons. In contrast, the shapes of the away-side jet distributions 
($\Delta \phi \geq \Delta \phi_{\rm min}$, AS) are qualitatively similar for associated 
mesons and baryons. For the central and mid-central collisions shown, these 
distributions are also broad and decidedly non-Gaussian, with evidence 
for local minima at $\Delta\phi=\pi$ ~\cite{Adler:2005ee}. 
They provide confirmation that the topological signatures for strong jet modification 
are reflected in the jet pair distributions for both associated baryons and 
mesons~\cite{contrast_ppg72}. 
The latter finding for baryons and mesons is an important constraint for models of 
strong jet-modification~\cite{Greco:2003mm,Armesto:2004,Stoecker:2005,Solana:2004}. 

	The integral of the extracted J($\Delta \phi$) distribution is the fraction of 
particle pairs associated with the jet, i.e., the jet pair fraction (JPF);
$JPF_{NS,AS} = \sum_{i\in NS,AS}{J(\Delta\phi_i)}/\sum_i{C(\Delta\phi_i)}$~\cite{Ajitanand2005}. 
We use it to determine the conditional yield $\langle N^{M,B} \rangle/\langle N_{\rm trig} \rangle$, or 
efficiency corrected pairs per trigger~\cite{Ajitanand2005};
%
\begin{equation}
\frac{\langle N^{M,B} \rangle}{\langle N_{\rm trig} \rangle} = JPF \times \frac{\langle N_{d}^{M,B} 
\rangle}{ \langle N^{\rm trig}_{s}\rangle \times \langle N^{M,B}_{s}\rangle}
\times \langle N^{M,B}_{eff}\rangle, 
\label{Eq:CY_decomp}
\end{equation} 
where $\langle N_{d}^{M,B} \rangle$ is the average number of detected hadron-meson(baryon) 
pairs per event, $\langle N^{\rm trig}_{s}\rangle$ and $\langle N^{M,B}_{s}\rangle$ are the 
detected singles rates for hadrons, and mesons and baryons respectively, 
and $\langle N^{M,B}_{eff}\rangle$ are the efficiency corrected singles rates. 
The systematic error associated with the latter is $\sim 10$\%.
A further division by the $p_{\rm T}$ bin width gives the conditional yield 
$CY = \frac{1}{N_{\rm trig}} \frac{dN}{dp_{\rm T}}$.

 	The conditional yields, for near- and away-side jet-associated mesons 
and baryons, are shown as a function of $p_{\rm T,assoc}^{M,B}$ and 
collision centrality in Fig.~\ref{fig:yld}.
The yields for associated mesons (Figs.~\ref{fig:yld} (a,b))
indicate an essentially exponential decrease with increasing $p_{\rm T,assoc}^M$, for both 
the near- and away-side jets. A decrease in the slope parameter (``temperature increase"), 
from peripheral to central collisions, is also apparent.
For a fixed $p_{\rm T,assoc}^M$, these yields also show an increase from 
peripheral to central events, albeit with a stronger dependence for the away-side 
jet. This trend is incompatible with in-vacuum fragmentation, but could be due to 
jet-like contributions from correlated soft partons which recombine upon 
hadronization~\cite{Fries:2004hd,new_calc}.

	The conditional yields for associated baryons differ strongly from 
those for associated mesons	(cf. Figs.~\ref{fig:yld} (c,d)). 
That is, they do not show an exponential dependence on $p_{\rm T}^B$ over 
the measured range, and the yields for the away-side jet are substantially
larger than those for the near-side jet. Interestingly, hadron-baryon pairs are 
disfavored in the same-side jet, relative to the away-side jet. 
A similar observation has been reported for charge 
selected ($pp$ and $\bar{p}\bar{p}$) jet correlations~\cite{Adare:2006nn}.

	For a given $p_{\rm T,assoc}^B$, the near- and away-side 
conditional yields increase as the collisions become more central, i.e., this
trend is similar to that for the associated mesons. 
However, the baryon yields show a much stronger increase with centrality~\cite{contrast_ppg72}, 
as might be expected if correlated soft partons recombine and contribute 
to the away-side jet correlations~\cite{new_calc}.
%
\begin{figure}[t]
\includegraphics[width=1.0\linewidth]{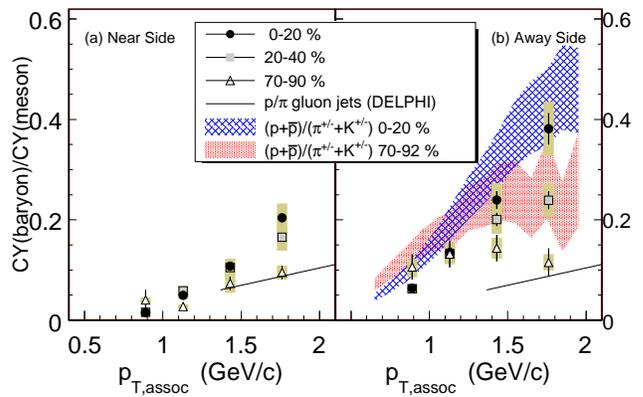}  
\caption{(Color online) Ratio of jet associated baryons to jet associated 
         mesons vs. $p_{\rm T,assoc}$ for the 0-20\%, 20-40\% and 70-90\% most 
         central collisions. The hatched bands indicate inclusive $B/M$ ratios (see text).
\label{fig:rat}}
\end{figure}

	The ratio of jet-associated baryons to jet-associated mesons is 
shown as a function of associated particle $p_{\rm T}$ in Fig.~\ref{fig:rat}; 
the left and right panels show the ratios for the near- and
away-side jets respectively, for three centrality selections 
as indicated. These ratios clearly increase with $p_{\rm T}$, and with 
centrality for $p_{\rm T} \agt 1.4$~GeV/$c$.
For peripheral collisions, the near side ratios compare well to the 
$p/\pi$ ratio, for jets produced in $e^{+}+e^{-}$ collisions 
(line)~\cite{Abreu:2000nw}. For more central collisions, the near- 
and away-side ratios are much larger, suggesting that the medium influences 
the relative composition of the associated particles. 

	The hatched bands in Fig.~\ref{fig:rat}(b), show the inclusive $B/M$ ratios 
(uncorrected for baryon and meson feed-down) as a function of $p_{\rm T}$ for the 
0-20\% and 70-92\% most central Au+Au collisions~\cite{Adler:2003cb}; 
an estimate of these ratios, after feed-down corrections, 
is within the systematic errors indicated by the bands. 
These ratios indicate that the trend of the centrality dependent baryon 
enhancement, apparent in the jet-associated conditional yields, is 
similar to that observed for the inclusive particle yields. They suggest that the 
underlying mechanism for baryon enhancement in both the inclusive and the current 
away-side jet measurements, have a common origin. Since recombination 
models 
can explain the enhancement of inclusive baryon 
yields, a qualitative explanation is that the away-side jet-like correlations 
result from the recombination of correlated soft partons induced via strong 
parton-medium interactions.   

	In summary, we have measured per-trigger yield distributions for 
jet-associated mesons and baryons over a wide range of centrality and $p_{\rm T}$ 
in Au+Au collisions. The distributions for both
species show similar shape modifications for the away-side jet, compatible 
with several jet modification models~\cite{Greco:2003mm,Armesto:2004,
Stoecker:2005,Solana:2004}. The conditional yield distributions for mesons and 
baryons show different dependencies on collision centrality  and associated particle 
$p_{\rm T}$. The ratio of jet-associated baryons to mesons 
increases with centrality and $p_{\rm T}$, similar to the data 
for inclusive measurements. 
These results can be qualitatively understood in terms of 
parton-medium interactions which induce correlations between 
soft partons, followed by recombination at hadronization~\cite{Fries:2004hd,new_calc}.
Future quantitative model comparisons are required to fully validate this mechanistic 
scenario. However, the current measurements offer important new insight that may  
reconcile the dichotomous observations that a simple quark recombination 
ansatz does not account for jet-induced hadron correlation measurements, 
but can explain the enhancement of baryon emission and the ``universal" elliptic flow 
of hadrons at intermediate $p_{\rm T}$.



We thank the staff of the Collider-Accelerator and Physics
Departments at BNL for their vital contributions.  
We thank P. Danielewicz and S. Pratt for their interest and input.
We acknowledge support from 
the Office of Nuclear Physics in DOE Office of Science and NSF (USA), 
MEXT and JSPS (Japan), 
CNPq and FAPESP (Brazil), 
NSFC (China), 
IN2P3/CNRS and CEA (France),
BMBF, DAAD, and AvH (Germany),
OTKA (Hungary),
DAE (India),
ISF (Israel), 
KRF and KOSEF (Korea), 
RMIST, RAS, and RMAE (Russia), 
VR and KAW (Sweden), 
US CRDF for the FSU, 
US-Hungarian NSF-OTKA-MTA, 
and US-Israel BSF.


\end{document}